\documentclass[pra,twocolumn,showpacs]{revtex4}
\usepackage{graphicx}
\usepackage{dcolumn}
\usepackage{bm}
\usepackage{amssymb}
\usepackage{amsmath}
\usepackage{epsfig}

\begin{document}

\title{Entanglement measurement based on two-particle interference}
\author{Jian-Ming Cai}
\email{jmcai@mail.ustc.edu.cn}
\author{Zheng-Wei Zhou}
\author{Guang-Can Guo}
\affiliation{Key Laboratory of Quantum Information, University of Science and Technology
of China, Chinese Academy of Sciences, Hefei, Anhui 230026, China}

\begin{abstract}
We propose a simple and realizable method using a two-particle
interferometer for the experimental measurement of pairwise entanglement,
assuming some prior knowledge about the quantum state. The basic idea is
that the properties of the density matrix can be revealed by the single- and
two-particle interference patterns. The scheme can easily be implemented
with polarized entangled photons.
\end{abstract}

\maketitle


\textbf{\textit{Introduction}} Quantum entanglement is a kind of
indispensable resource for exponential speedups of future quantum computers
\cite{Ekert&Jozsa,Nielsen & Chuang} and long distance secret quantum
communications \cite{Duan,Ekert,Gisin2002}. However, for practical physical
systems couplings with the external environment are unavoidable, which will
result in various decoherence processes. Therefore, entanglement will be
reduced consequentially \cite%
{Simon&Kempe,Dur&Briegel,Carvalho&Mintert&Buchleitner}, especially during
creation in real experiments with noise and imperfections and distribution
through a lossy channel. This makes it crucial to find efficient methods to
detect entanglement, not only to test whether a given state --- both pure
states and mixed states --- is entangled or not, but also to determine the
degree of entanglement.

In the recent years, several methods for measurement of entanglement have
been proposed. The most straightforward way is to reconstruct the quantum
state fully through quantum tomography \cite{White}. For systems of two
qubits, it requires nine different measurement settings to determine 15
parameters which describe a general two-qubit state. However, not all these
parameters are necessary for measurement of entanglement, assuming some
prior knowledge of the density matrix. In \cite{Horodecki&Ekert}, P.
Horodecki and A. Ekert proposed a direct and efficient scheme which provides
the estimation of the degree of entanglement of an unknown quantum state.
This scheme is based on the quantum network which can evaluate certain
nonlinear functionals of density matrixes. However, it requires the
implementation of control unitary operations, which may not be accomplished
very easily in real experiments. Another different approach is the method of
entanglement witness operator $W$ \cite{Peres,Horodecki}. For a density
matrix $\rho$, which is entangled, the expectation value is negative $%
Tr(W\rho)<0$, nevertheless the expectation value is positive $%
Tr(W\rho_{sep})\geq 0$ for all separable states. With a few local
measurements \cite{Sanpera et.al}, one can obtain the expectation value of
the entanglement witness and then detect the presence of entanglement. In
the first experiment realization of entanglement witness \cite{Barbieri},
three different measurement settings are required for Werner states \cite%
{Sanpera et.al}.

In this paper, we introduce a new method for experimental measurement of
entanglement using a two-particle interferometer, which is extended from
standard one-particle interferometry. Lots of theoretical analyses and
experiments have been carried out in this field \cite%
{Two-Interference,Kaszlikowski,Zeilinger}. Complementarity of one-particle
and two-particle interference \cite{Jaeger1993,Jaeger1995} has been
revealed. Here we reexamine the properties of one- and two-particle
interference patterns of two particles in entangled states. It is showed
that, assuming some prior knowledge about the entangled states, the degree
of entanglement can be reflected by the properties of the interference
patterns. Therefore, the degree of entanglement can be measured by study
single- and joint-detection probabilities through a two-particle
interferometer. This presents a new kind of utility of quantum interference
in quantum information processing. Since techniques of two-particle
interferometry have been developed very well, thus our scheme for
measurement of entanglement is very simple and can be easily realized in
practical experiments.

\textbf{\textit{Two-particle interferometer}} A schematic two-particle
interferometer \cite{Jaeger1993,Jaeger1995} is depicted in Fig. 1. The
source $S$ produce a pair of particles $1$ and $2$, which can be realized
through a laser-pumped down-converting crystal \cite{Zeilinger}. Particle $1$
propagates along paths $A$ and/or $A^{\prime}$, passing through the passive
lossless transducer $T_{1}$, and then emerges out in either $U_{1}$ or $%
L_{1} $. Similarly, particle $2$ propagates along paths $B$ and/or $%
B^{\prime}$, passing through the passive lossless transducer $T_{2}$, and
then emerges out in either $U_{2}$ or $L_{2}$.

\begin{figure}[htb]
\epsfig{file=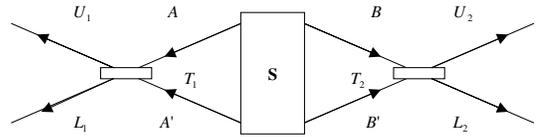,width=7cm}
\caption{Schematic two-particle interferometer. Two particles emit from the
source $S$ and pass through the passive lossless transducers $T_{1}$ and $%
T_{2}$. Then they are detected at the outports $U_{1}$, $L_{1}$ and $U_{2}$,
$L_{2}$ respectively.}
\end{figure}

For simplicity, we denote $|0\rangle_{1}=|A\rangle$, $|1\rangle_{1}=|A^{%
\prime}\rangle$ and $|0\rangle_{2}=|B\rangle$, $|1\rangle_{2}=|B^{\prime}%
\rangle$ in the rest of this paper. Likewise, for the output Hilbert space $%
|0_{o}\rangle_{i}=|U_{i}\rangle$ and $|1_{o}\rangle_{i}=|L_{i}\rangle$ $(
i=1,2 )$. The general entangled two-particle state $\rho$ generated from the
source $S$ is in the Hilbert space spanned by the basis $\{|0\rangle_{1}|0%
\rangle_{2}, |0\rangle_{1}|1\rangle_{2}, |1\rangle_{1}|0\rangle_{2},
|1\rangle_{1}|1\rangle_{2}\}$. The transducer $T_{1}$ implements the unitary
unimodular map for particle $1$ as follows \cite{Jaeger1995}
\begin{eqnarray}
T_{1}|0^{\prime}\rangle_{1}&=&ae^{i\phi_{a}}|0_{o}\rangle_{1}+be^{i%
\phi_{b}}|1_{o}\rangle_{1}  \notag \\
T_{1}|1^{\prime}\rangle_{1}&=&-be^{-i\phi_{b}}|0_{o}\rangle_{1}+ae^{-i%
\phi_{a}}|1_{o}\rangle_{1}
\end{eqnarray}
where $\{{|0^{\prime}\rangle_{1}, |1^{\prime}\rangle_{1}}\}$ is an
orthonormal basis for the space spanned by $|0\rangle_{1}$ and $%
|1\rangle_{1} $, $a$ and $b$ are real numbers and satisfy $a^2+b^2=1$.
Similarly, the transducer $T_{2}$ implements the unitary unimodular map for
particle $2$ as \cite{Jaeger1995}
\begin{eqnarray}
T_{2}|0^{\prime}\rangle_{2}&=&ce^{i\phi_{c}}|0_{o}\rangle_{2}+de^{i%
\phi_{d}}|1_{o}\rangle_{2}  \notag \\
T_{2}|1^{\prime}\rangle_{2}&=&-de^{-i\phi_{d}}|0_{o}\rangle_{2}+ce^{-i%
\phi_{c}}|1_{o}\rangle_{2}
\end{eqnarray}
where $\{{|0^{\prime}\rangle_{2}, |1^{\prime}\rangle_{2}}\}$ is an
orthonormal basis for the space spanned by $|0\rangle_{2}$ and $%
|1\rangle_{2} $, $c$ and $b$ are real numbers and satisfy $c^2+d^2=1$. Then
after passing through two transducers $T_{1}$ and $T_{2}$, the state of
particle $1$ and particle $2$ is transformed into $\rho
^{\prime}=(T_{1}\bigotimes T_{2})\rho (T_{1}\bigotimes T_{2})^\dagger$. Here
$\rho ^{\prime}$ is in the Hilbert space spanned by the basis $%
\{|0_{o}\rangle_{1}|0_{o}\rangle_{2}, |0_{o}\rangle_{1}|1_{o}\rangle_{2},
|1_{o}\rangle_{1}|0_{o}\rangle_{2}, |1_{o}\rangle_{1}|1_{o}\rangle_{2}\}$.
Now we can measure the single detection probability $P(U_{1}) =\langle
0_{o}|_{2}\langle
0_{o}|_{1}\rho^{\prime}|0_{o}\rangle_{1}|0_{o}\rangle_{2}+\langle
1_{o}|_{2}\langle 0_{o}|_{1}\rho^{\prime}|0_{o}\rangle_{1}|1_{o}\rangle_{2}$
and the probability of joint output $P(U_{1}U_{2}) = \langle
0_{o}|_{2}\langle 0_{o}|_{1}\rho^{\prime}|0_{o}\rangle_{1}|0_{o}\rangle_{2}$%
. Other analogous probabilities $P(U_{1}L_{2})$, $P(L_{1} U_{2})$, $%
P(L_{1}L_{2})$ and $P(U_{2})$, $P(L_{1})$, $P(L_{2})$ can be written in a
similar way.

\textbf{\textit{One-particle interference}} We start by investigating the
single-particle fringe visibility. Suppose particle $1$ is prepared in an
arbitrary two-level quantum state $\rho$. For our purpose $\rho$ is
expressed in the basis $\{|0\rangle_{1}, |1\rangle_{1}\}$
\begin{equation}
\rho=\left(
\begin{array}{cccc}
\rho_{00} & \rho_{01} &  &  \\
\rho_{10} & \rho_{11} &  &
\end{array}
\right)
\end{equation}
When particle $1$ enters and exits the transducer $T_{1}$, it experiences
the unitary map induced by $T_{1}$ and its state changes into $%
\rho^{\prime}=T_{1}\rho T^\dagger_{1}$ consequently. We express $%
\rho^{\prime}$ in the basis $\{|0_{o}\rangle_{1}, |1_{o}\rangle_{1}\}$ and
then the single detection probability can be obtained straightforwardly $%
P(U_{1}) = \rho^{\prime}_{0_{o}0_{o}}$, that is
\begin{equation}
P(U_{1}) =
a^2\rho_{00}+b^2\rho_{11}-abe^{i\phi_{1}}\rho_{01}-abe^{-i\phi_{1}}\rho_{10}
\end{equation}
where $\phi_{1}=\phi_{a}+\phi_{b}$. We now turn to the single-particle
fringe visibility $V_{1}$, which is defined as follows \cite{Jaeger1993}
\begin{equation}
V_{1}=\frac{[P(U_{1})]_{max}-[P(U_{1})]_{min}}{%
[P(U_{1})]_{max}+[P(U_{1})]_{min}}
\end{equation}
To find the extreme of $P(U_{1})$ we first note that $P(U_{1}) =
a^2\rho_{00}+b^2\rho_{11}-2abRe[\rho_{01}e^{i\phi_{1}}]$. Then for some
fixed $a$ and $b$, the extreme are achieved $[P(U_{1})]_{max}=a^2%
\rho_{00}+b^2\rho_{11}+2|ab\rho_{01}|$ and $[P(U_{1})]_{in}=a^2\rho_{00}+b^2%
\rho_{11}-2|ab\rho_{01}|$ by choosing appropriate phase parameters $\phi_{a}$
and $\phi_{b}$. Assuming $|a|=\cos{(\mu/2)}$ and $|b|=\sin{(\mu/2)}$ with $%
0\leq\mu\leq\pi$, we can get
\begin{eqnarray}
[P(U_{1})]_{max} &=&\frac{1}{2}+\frac{\rho_{00}-\rho_{11}}{2}\cos{\mu}%
+|\rho_{01}|\sin{\mu}  \notag \\
&=&\frac{1}{2}+\frac{\sqrt{I_{1}}}{2}\sin{(\mu+\mu^{\prime})}
\end{eqnarray}
where $I_{1}=(\rho_{00}-\rho_{11})^2+4|\rho_{01}|^2$ and $\mu^{\prime}$
satisfies that $\sin{\mu^{\prime}}=(\rho_{00}-\rho_{11})/\sqrt{I_{1}}$ and $%
\cos{\mu^{\prime}}=2|\rho_{01}|/\sqrt{I_{1}}$. Therefore, it can be easily
seen that
\begin{equation}
[P(U_{1})]_{max}=\frac{1+\sqrt{I_{1}}}{2}
\end{equation}
The minimum value $[P(U_{1})]_{min}$ can be obtained in a similar way
\begin{equation}
[P(U_{1})]_{min}=\frac{1-\sqrt{I_{1}}}{2}
\end{equation}
Now according to the definition in Eq. (5), the square of the
single-particle fringe visibility for particle $1$ is
\begin{equation}
V^{2}_{1}=I_{1}
\end{equation}

Before proceeding to the next part, it is worth for us to discuss the
physical meaning of the above result in Eq. (9). We find that $%
I_{1}=\rho^2_{00}+\rho^2_{11}-2\rho_{00}\rho_{11}+4|\rho_{01}|^2=2(%
\rho^2_{00}+\rho^2_{11}+2|\rho_{01}|^2)-1$. Here we have used $%
\rho_{00}+\rho_{11}=1$. Note that $Tr\rho^2=\rho^2_{00}+\rho^2_{11}+2|%
\rho_{01}|^2$, therefore we have $I_{1}=2Tr\rho^2-1$, which is just the
total information content in a two-state quantum system \cite%
{Brukner&Zeilinger}. The total information is defined as the sum over a
complete set of mutually complementary measurements. For a spin-1/2 particle
in the state $\rho$, a complete set of mutually complementary experiments
consists of three measurements. The relation, depicted above, shows that the
single-particle fringe visibility $V_{1}$ reflects the total information
content in the quantum state $\rho$. Conversely, through one-particle
interferometry the total information content and then the purity of the
quantum state can be determined too.

\textbf{\textit{Scheme for measurement of entanglement}} In the following we
consider a special class of entangled two-particle states of the form
\begin{equation}
\rho=p|\psi\rangle\langle \psi | +(1-p)\frac{I}{4}, \;\;\;\;0\leq p\leq 1
\end{equation}
where $|\psi\rangle$ is an arbitrary pure state $|\psi \rangle =
\lambda_{1}|0\rangle_{1}|0\rangle_{2} +\lambda_{2}|0\rangle_{1}|1\rangle_{2}
+\lambda_{3}|1\rangle_{1}|0\rangle_{2}
+\lambda_{4}|1\rangle_{1}|1\rangle_{2} $ with the normalization condition $%
|\lambda_{1}|^2 +|\lambda_{2}|^2 + |\lambda_{3}|^2 +|\lambda_{4}|^2=1$.
According to the Schmidt decomposition theorem, $|\psi\rangle$ can be
expressed as $|\psi\rangle=\alpha
|0^{\prime}\rangle_{1}|1^{\prime}\rangle_{2}+\beta
|1^{\prime}\rangle_{1}|0^{\prime}\rangle_{2}$, where $|\alpha|^2+|\beta|^2=1$%
. Without loss of generality, we can suppose $|\alpha|^2\geq|\beta|^2$ here.
When $|\psi\rangle=|\Psi_{-}\rangle$ the above entangled states is just the
Werner states \cite{Werner}. The state $\rho$ given in Eq. (10) exists in
many practical realizations. The pure part $|\psi \rangle \langle \psi|$ is
the ideal entangled state to be produced through real experiments with noise
and imperfections or to be distributed between two distant parties over a
lossy channel. And the maximally chaotic part $I/4$ is induced by the
decoherence processes. When $p$ is larger than some critical value, the
above state $\rho$ is entangled. Otherwise, it is separable. The degree of
entanglement is not only dependent on $p$ but also determined by the pure
state $|\psi\rangle$.

In the basis $\{|0^{\prime}\rangle_{1}|0^{\prime}\rangle_{2},
|0^{\prime}\rangle_{1}|1^{\prime}\rangle_{2},
|1^{\prime}\rangle_{1}|0^{\prime}\rangle_{2},
|1^{\prime}\rangle_{1}|1^{\prime}\rangle_{2}\}$ the state $\rho$ can be
expressed as
\begin{equation}
\rho =\left(
\begin{array}{cccc}
w & 0 & 0 & 0 \\
0 & x & z & 0 \\
0 & z^{\ast } & y & 0 \\
0 & 0 & 0 & w%
\end{array}
\right)
\end{equation}
where $w=(1-p)/4$, $x=(1-p)/4+p|\alpha|^2$, $y=(1-p)/4+p|\beta|^2$ and $%
z=p\alpha\beta^{\ast}$. We denote the time-reversed matrix of $\rho$ as $%
\widetilde{\rho }=(\sigma _{y}\otimes \sigma _{y})\rho ^{\ast }(\sigma
_{y}\otimes \sigma _{y})$. After straightforward calculations, the square
roots of four eigenvalues of $\rho \widetilde{\rho }$ can be expressed as $%
\{w, w, \sqrt{xy}-|z|, \sqrt{xy}+|z|\}$. Therefore, the entanglement of $%
\rho $, measured by concurrence \cite{ConC}, is
\begin{equation}
C=2\max{\{p|\alpha\beta|-\frac{1-p}{4},0\}}
\end{equation}
Our main goal is to measure $C$ from the fringes of single- and two-particle
interference. The visibility of single-particle fringe has been analyzed
above. In order to achieve the goal of entanglement measurement, we need to
investigate the property of two-particle fringe.

In the two-particle interferometer, after the two particles passing through
the passive and lossless transducers $T_{1}$ and $T_{2}$, the entangled
state $\rho $ changes into $\rho ^{\prime }=(T_{1}\bigotimes T_{2})\rho
(T_{1}\bigotimes T_{2})^{\dagger }$. Therefore, the probability of joint
detection by ideals detectors placed at the output forts $U_{1}$ and $U_{2}$
is $P(U_{1}U_{2})=\langle 0_{o}|_{2}\langle 0_{o}|_{1}\rho ^{\prime
}|0_{o}\rangle _{1}|0_{o}\rangle _{2}$, which can be calculated as follows
\begin{eqnarray}
P(U_{1}U_{2}) &=&a^{2}(c^{2}w+d^{2}x)+b^{2}(c^{2}y+d^{2}w)  \notag \\
&&+abcd(p\alpha \beta ^{\ast }e^{i\phi }+p\alpha ^{\ast }\beta e^{-i\phi })
\end{eqnarray}%
where $a$, $b$, $c$, $d$ and $\phi =\phi _{a}+\phi _{b}-\phi _{c}-\phi _{d}$
are parameters describing the transducers $T_{1}$ and $T_{2}$. Note that $%
P(U_{1}U_{2})=a^{2}(c^{2}w+d^{2}x)+b^{2}(c^{2}y+d^{2}w)+2pabcdRe[\alpha
\beta ^{\ast }e^{i\phi }]$. Then when the parameters $a$, $b$, $c$, $d$ are
fixed, the maximum probability of joint output are achieved $%
[P(U_{1}U_{2})]_{max}=a^{2}(c^{2}w+d^{2}x)+b^{2}(c^{2}y+d^{2}w)+2pabcd|%
\alpha \beta |$ by choosing appropriate phase parameters $\phi _{a}$, $\phi
_{b}$ and $\phi _{c}$, $\phi _{d}$. Assuming $|a|=\cos {(\mu /2)}$, $%
|b|=\sin {(\mu /2)}$ and $|c|=\cos {(\upsilon /2)}$, $|d|=\sin {(\upsilon /2)%
}$ with $0\leq \mu ,:\upsilon \leq \pi $, we can get
\begin{eqnarray}
\lbrack P(U_{1}U_{2})]_{max} &=&\frac{1}{4}+m(\cos {\mu }-\cos \upsilon )-%
\frac{p}{4}\cos {\mu }\cos \upsilon   \notag \\
&&+n\sin {\mu }\sin \upsilon
\end{eqnarray}%
where $m=p(|\alpha |^{2}-|\beta |^{2})/4,n=p|\alpha \beta |/2$.

Since $|\alpha \beta |\leq (|\alpha |^{2}+|\beta |^{2})/2=1/2$, thus $n\leq
p/4$. Therefore it follows that
\begin{eqnarray}
\lbrack P(U_{1}U_{2})]_{max} &=&\frac{1}{4}+m(\cos {\mu }-\cos \upsilon )-%
\frac{p}{4}\cos {(\mu +\upsilon )}  \notag \\
&-&(\frac{p}{4}-n)\sin {\mu }\sin \upsilon  \leq \frac{p+1}{4}+2m  \notag \\
&=&\frac{p+1}{4}+\frac{p}{2}(|\alpha |^{2}-|\beta |^{2})
\end{eqnarray}%
The equation is satisfied when $\mu =0$ and $\upsilon =\pi $. For the sake
of discussion later, we denote $P_{12}=[P(U_{1}U_{2})]_{max}=(p+1)/4+p(|%
\alpha |^{2}-|\beta |^{2})/2$.

In addition, for the state $\rho$ given in Eq. (10), the single-particle
fringe visibility, according to Eq. (9), is $V^{2}_{1}=I_{1}=2Tr\rho^{2}_{1}$%
. Here $\rho_{1}=Tr_{2}\rho$ is the reduced density matrix of particle $1$.
Therefore, the single-particle fringe visibility can be written
straightforwardly
\begin{equation}
V^{2}_{1}=p^2(1-4|\alpha|^2|\beta|^2)
\end{equation}

Note that $|\alpha|^2+|\beta|^2=1$, from Eq. (15) and (16) we can get
\begin{equation}
P_{12}= (p+1)/4+V_{1}/2
\end{equation}

Therefore, the parameter $p$ can be evaluate as
\begin{equation}
p=4P_{12}-2V_{1}-1
\end{equation}

By Eq. (12), (18) and note that $2p|\alpha \beta |=\sqrt{p^{2}-V_{1}^{2}}$,
the amount of entanglement in the state $\rho $ is
\begin{eqnarray}
C &=&\max {\{2P_{12}-V_{1}-1}  \notag \\
&&{+\sqrt{(4P_{12}-3V_{1}-1)(4P_{12}-V_{1}-1)},0\}}
\end{eqnarray}%
It can be seen that in order to determine the amount of entanglement in the
state $\rho $ give in Eq. (12), we only need to measure the single-particle
fringe visibility $V_{1}$ and the maximum probability of joint output $%
P_{12} $. And then based on Eq. (19), the value $2P_{12}-V_{1}-1+\sqrt{%
(4P_{12}-3V_{1}-1)(4P_{12}-V_{1}-1)}$ can be evaluated. If this value is
negative then the state $\rho $ is separable. Otherwise, this value is just
the concurrence of $\rho $ to be measured. That is we have proposed a simple
method based on two-particle interference for measurement of entanglement,
assuming some prior knowledge about the quantum state. For the Werner class
of mixed states, only two quantities $V_{1}$ and $P_{12}$ are required to be
measured in application of our scheme.

\textbf{\textit{Discussions}} From the concurrence expression in Eq. (19),
we can see a natural requirement for the single-particle fringe visibility $%
V_{1}$ and the maximum probability of joint output $P_{12}$ that is $%
4P_{12}-3V_{1}-1 \geq 0$. From Eq. (17), we can see that this requires that $%
4P_{12}-3V_{1}-1 = p-V_{1}\geq 0$. Obviously, the above condition can be
satisfied according to the expression for the single-particle fringe
visibility $V_{1}$ in Eq. (16). It should be pointed out that the above
requirement $4P_{12}-3V_{1}-1 \geq 0$ is satisfied for the Werner class of
mixed states given in Eq. (10). However, it is not an universal requirement
for any general two-particle states.

An illustrative demonstration of our scheme for entanglement measurement can
be implemented with entangled polarized photons in Werner states $%
\rho=p|\Psi_{-}\rangle \langle \Psi_{-}| +(1-p)\frac{I}{4}$, $0\leq p\leq 1$%
, produced via spontaneous parametric down-conversion (SPDC) \cite%
{Kwiat,Guo,Barbieri@2004}. The combination of two half-wave plate (HWP) and
one quarter-wave plate (QWP) yields an arbitrary unitary rotation on
polarized photons \cite{James}, which functions as the passive lossless
transducers $T_{1}$ and $T_{2}$. Together with some linear optical
instruments, such as polarizing beam splitter (PBS), the single-particle
fringe visibility $V_{1}$ and the maximum joint probability $P_{12}$ can be
obtained straightforwardly. Therefore, the degree of entanglement in the
Werner states can be measured successfully based on our scheme.

The idea of this paper can be easily applied for the detection of
entanglement in some other special class of quantum states other than the
Werner class of mixed states given in Eq. (10). For example, the initial
states for entanglement purification \cite{pan} and the Gisin mixed states
\cite{Peres,Gisin}. In fact, the interferometric method of entanglement
measurement is applicable to those mixed states that need only two
independent parameters to characterize the degree of entanglement. For
example, these two independent parameters for Werner states are $p$ and $%
\alpha$, for Gisin mixed states are $a$ and $x$ in \cite{Peres,Gisin}. The
measure of the single-particle fringe visibility $V_{1}$ and the maximum
joint probability $P_{12}$ will present two independent equations for these
two independent parameters. Therefore, the degree of entanglement can be
measured through the interferometric method. Although, we need some prior
knowledge about the state and the relation between entanglement and $V_{1}$,
$P_{12}$ maybe different for different class of mixed states. However, our
scheme is suitable for many practical situations \cite{Barbieri@2004,Guo,pan}%
. Furthermore, we can generalize our scheme to measure the true $3-$qubit
entanglement \cite{CKW} of $3-$qubit pure states. Similarly, only
two-particle interferometer is required, which will be discussed detailedly
in our future work. In addition, whether we can make use of multi-particle
interferometry to measure the degree of entanglement in general multi-qubit
states is still an open and interesting question.

\textbf{\textit{Conclusions}} In summary, we analyze the relation between
the interference patterns and the properties of the quantum states. It is
shown that the single-particle fringe visibility is just related to the
information content in the quantum state of the particle. This result also
links the purity of the state with the single-particle fringe visibility. In
addition, we propose a simple scheme for measurement of entanglement using a
two-particle interferometer, assuming some prior knowledge about the quantum
state. The scheme is applicable for several special class of mixed states,
e.g Werner class of mixed states and Gisin mixed states. An illustrative
demonstration of our scheme can be easily implemented with polarized photon
pairs produced via spontaneous parametric down-conversion. The optical
instruments and quantum interferometric techniques required for practical
experiments are all achievable.

\textbf{\textit{Acknowledgments}} We thank Dr. B. H. Liu for helpful
discussions. This work was funded by National Fundamental Research Program
(2001CB309300) and NCET-04-0587, the Innovation funds from Chinese Academy
of Sciences, and National Natural Science Foundation of China (Grant No.
60121503, 10574126).

\end{document}